\documentclass[aps,prl,twocolumn,showpacs]{revtex4}
\usepackage{amssymb}
\usepackage{amsmath, amsthm}
\newcommand{\be}{\begin{equation}}
\newcommand{\ee}{\end{equation}}
\newcommand{\bc}{\begin{center}}
\newcommand{\ec}{\end{center}}

\begin{document}


\title{Cosmological apparent and trapping horizons} 


\author{Valerio Faraoni}
\email[]{vfaraoni@ubishops.ca}
\affiliation{Physics Department and STAR Research Cluster, 
Bishop's University\\
2600 College Street, Sherbrooke, Qu\'ebec, Canada J1M~1Z7
}

\begin{abstract} 
The dynamics of particle, event, and 
apparent horizons in FLRW space are discussed. The apparent 
horizon is trapping when the  Ricci curvature  is positive.  
This simple criterion coincides with the condition for the 
Kodama-Hayward apparent horizon temperature 
to be positive, and also discriminates between timelike and 
spacelike character of the apparent horizon. We discuss also the 
entropy of apparent cosmological horizons in extended theories of 
gravity and we use the generalized 2nd law to discard an exact 
solution of Brans-Dicke gravity as unphysical.
\end{abstract}



\maketitle

\section{Introduction}

Black hole thermodynamics \cite{Waldthermo} links classical 
gravity and quantum mechanics and constitutes a major advancement 
of the theoretical physics of the 1970's. The discovery by 
Bekenstein \cite{Bekenstein} and Hawking 
\cite{Hawkingradiation1, Hawkingradiation2} that 
black hole horizons have entropy and temperature associated with 
them allowed for the formulation of a complete thermodynamics of 
black holes. It is widely believed that formulating also a 
statistical mechanics to explain black hole 
thermodynamics in terms of microscopic degrees of freedom 
requires a fully developed theory of quantum gravity, which is 
not yet available. 

Soon after the discovery of Hawking radiation 
\cite{Hawkingradiation1, Hawkingradiation2}, 
Gibbons and Hawking discovered that also the de Sitter 
cosmological event horizon is endowed with a temperature and an 
entropy, similar to the Schwarzschild horizon 
\cite{GibbonsHawking77}. Later, it was 
realized that the notion of black hole event horizon, which 
requires one to know the entire future development and causal 
structure of spacetime, is essentially useless for practical 
purposes. The teleological event horizon is not an easy quantity 
to compute: this feature has been emphasized by the development 
of numerical relativity. In the sophisticated simulations of 
black hole collapse available nowadays, outermost marginally 
trappes surfaces and {\em apparent} horizons are used as 
proxies for event 
horizons \cite{BaumgarteShapiro03}. While the early literature on 
black holes and the development of black hole thermodynamics in 
the 1970's focused on static and stationary black holes, for 
which apparent and event horizons coincide, dynamical situations 
such as the intermediate stages of black hole collapse, black 
hole evaporation backreacting on its source, and black holes 
interacting with non-trivial environments ({\em e.g.}, with 
another black hole or compact object, or with a cosmological 
background)  require the generalization of the concept of event 
horizon to situations in which no timelike Killing vector is 
available. For this purpose, the concepts of apparent, trapping, 
isolated, dynamical, and slowly evolving horizons were developed 
(see \cite{Boothreview, Nielsenreview, AshtekarKrishnan} for 
reviews). 

In addition to black hole horizons, also cosmological 
horizons have been the subject of intense scrutiny. The de Sitter 
event horizon considered by Gibbons and Hawking as a 
thermodynamical system \cite{GibbonsHawking77} is  
static due to the high symmetry of de Sitter space, which admits 
a timelike Killing vector, and plays a role analogous to that 
of the Schwarzschild event horizon among black holes. For 
more general Friedmann-Lemaitre-Robertson-Walker (FLRW) spaces, 
which do not admit such a  Killing vector, the particle and 
event horizons are familiar from standard cosmology textbooks. 
However, they do not exist in all FLRW spaces and they  do not 
seem suitable for formulating consistent thermodynamics 
(\cite{Collins92, HaywardMukohyamaAshworth99PLA, Hayward98CQG, 
BakRey00CQG, Bousso05PRD, NielsenYeom09} and references 
therein).  Instead, the 
FLRW {\em apparent} horizon, which always exists contrary to the 
event and particle horizons, seems a better candidate. In this 
paper we 
reconsider our knowledge of this horizon and try to deepen our 
understanding of it. Specifically, we derive a simple criterion 
for the apparent horizon to be also a trapping horizon and we 
show that the Kodama-Hayward temperature, which is based on the 
Kodama vector playing the role of the timelike Killing vector 
outside the horizon, is positive if and only if  the apparent 
horizon is trapping. The causal character of this surface is 
related to this criterion.

The thermodynamics of the FLRW apparent horizon has seen much  
interest recently, with many authors deriving the temperature of 
this horizon with the Hamilton-Jacobi variant of the 
Parikh-Wilczek ``tunneling'' approach \cite{ParikhWilczek00}. 
However, different definitions of surface gravity can be applied 
to this calculation in order to define the energy of (scalar) 
particles, corresponding to a background notion of time, and 
these different  prescriptions provide different notions of 
temperature. The Kodama-Hayward prescription seems to stand out 
among its competitors because the Kodama vector is associated 
with a  conserved current even in the absence of a timelike 
Killing  vector \cite{Kodama}, a fact called the ``Kodama 
miracle''  \cite{AbreuVisser10}  which leads to several 
interesting 
results.  What is more, the Noether charge associated with the 
Kodama vector is the Misner-Sharp-Hernandez mass 
\cite{MisnerSharp, MisnerHernandez}, which is 
almost universally  adopted as the internal energy $U$ in horizon 
thermodynamics.  The Misner-Sharp-Hernandez mass, defined in 
spherical  symmetry, coincides with the Hawking 
quasi-local energy \cite{Hayward96PRD53} and, if we insist in 
using it in thermodynamics, the use of the Kodama-Hayward surface 
gravity follows naturally 
(although this point may be considered debatable by some).

In the next section, we review background material  
while deriving new formulas useful in the study of 
apparent and trapping horizons. Sec.~3 
discusses the various notions of horizons in 
FLRW space and their dynamics, and elucidates the causal 
character of the apparent horizon. The following section raises  
a 
question neglected in the literature, namely the condition under 
which  the apparent horizon is also a trapping horizon.  The 
simple criterion is  that the Ricci scalar must be positive. 
We then show that the Kodama-Hayward temperature is positive when 
the apparent horizon is trapping. Secs.~5 and~6 contain  
discussions of 
the thermodynamics of cosmological horizons in General 
Relativity (GR) and in extended 
theories 
of gravity  and uses the generalized 2nd law to reject as 
unphysical an exact solution of Brans Dicke theory. Sec.~7 
contains the  conclusions. We follow the notations of 
\cite{Waldbook}. The speed of light $c$, reduced Planck constant 
$\hbar$, and Boltzmann constant $K_B$ are set equal to unity, 
however they are occasionally restored for better clarity.

\section{Background}

The FLRW line element in comoving coordinates $\left( t,r, 
\theta, \varphi \right)$  is 
\be \label{FLRWlineelement}
ds^2=-dt^2+a^2(t) \left( \frac{dr^2}{1-kr^2} +r^2 
d\Omega_{(2)}^2 \right) 
\ee
where $k$ is the curvature index, $a(t)$ is the scale factor, 
and $d\Omega_{(2)}^2=d\theta^2+\sin^2\theta \ d\varphi^2$ is the 
line element on the unit 2-sphere. Sometimes different 
coordinates employing the areal radius  $R(t,r) \equiv 
a(t)r$ are useful.  Such coordinate systems 
include the pseudo-Painlev\'e-Gullstrand  and Schwarzschild-like 
coordinates,  which we introduce here for a general FLRW space.

Begin from the metric~(\ref{FLRWlineelement}); using 
the areal radius $R$, this line element assumes the   
pseudo-Painlev\'e-Gullstrand form
\begin{eqnarray} 
ds^2 & = & - \left( 1- \frac{H^2R^2}{1-kR^2/a^2} \right)dt^2 
-\frac{2HR}{1-kR^2/a^2} \, dtdR \nonumber\\
&&\nonumber\\
&{} & +\frac{dR^2}{1-kR^2/a^2}  +R^2 d\Omega_{(2)}^2  \,,
\label{FLRWPG}
\end{eqnarray}
where $H\equiv \dot{a}/a$ is the Hubble parameter and an overdot 
denotes differentiation with respect to the comoving time $t$. We 
use the word ``pseudo'' because the coefficient of $dR^2$ is 
not unity, as  required for Painlev\'e-Gullstrand  
coordinates \footnote{The literature contains ambiguous 
terminology for general FLRW spaces ({\em e.g.}, 
\cite{CaiCaoHu09}),  while the de Sitter case does not lend 
itself to these ambiguities \cite{Parikh02, Medved02}.} and the 
spacelike surfaces 
$t=$constant are not flat 
(unless $k=0$), which is regarded as the essential property of 
Painlev\'e-Gullstrand coordinates \cite{MartelPoisson01}. 

To transform to the Schwarzschild-like form, one first introduces 
the new time  $T$ defined by
\be
dT=\frac{1}{F} \left( dt +\beta dR \right) \,,
\ee
where $F$ is a (generally non-unique) integrating factor 
satisfying
\be
\frac{\partial}{\partial R}\left( \frac{1}{F} \right)=
\frac{\partial}{\partial t}\left( \frac{ \beta}{F} \right)
\ee
to guarantee that $dT$ is a locally exact differential, while 
$\beta(t,R)$ is a function to be determined. Substituting 
$ dt=FdT-\beta dR $ into the line element, one obtains
\begin{eqnarray}
ds^2 =& - & \left( 1-\frac{H^2R^2}{1-kr^2}\right) F^2dT^2 
\nonumber\\
&&\nonumber\\
&+ &  \left[ -\left( 1-\frac{H^2R^2}{1-kr^2} \right) \beta^2 
+\frac{2HR \beta + 1}{1-kr^2} \right] 
dR^2\nonumber\\
&&\nonumber\\
& + & 2 \left( 1-\frac{H^2R^2}{1-kr^2} \right)F \beta dTdR 
\nonumber\\
&&\nonumber\\
&-& \frac{2HR F}{1-kr^2}\, dTdR+R^2 d\Omega_{(2)}^2 \,.
\end{eqnarray}
By choosing
\be
\beta  = \frac{HR}{ 1-H^2R^2 -kr^2} \,, 
\ee
the cross-term proportional to $dTdR$ is eliminated and one 
obtains the FLRW line element in the Schwarzschild-like form  
\footnote{For de Sitter space with $H=$const., it is possible to 
set $F\equiv 1$ and the metric is cast in static coordinates, 
which cover the region $0\leq R\leq H^{-1}$.} 
\begin{eqnarray}
ds^2 & = & -\left( 1- \frac{H^2R^2}{1-kR^2/a^2} \right) F^2dT^2 
\nonumber\\
&&\nonumber\\
&+&  \frac{dR^2}{1-kR^2/a^2 -H^2R^2} 
 +  R^2d\Omega_{(2)}^2 \,,  \label{FLRWSchwarzschild}
\end{eqnarray}
where $F=F(T, R), a$, and $H$ are implicit functions of 
$T$. 

Horizons in spherical symmetry are discussed in a clear and 
elegant way by Nielsen and  Visser in \cite{NielsenVisser06} 
(see also \cite{NielsenYeom09}).  These authors consider the most 
general spherically symmetric metric  with a spherically 
symmetric spacetime slicing, which assumes the 
form (in Schwarzschild-like coordinates)
\begin{eqnarray} 
ds^2 & = & -\mbox{e}^{-2\phi (t, R)} \left[ 1-\frac{2M(t,R)}{R} 
\right] dt^2 +\frac{dR^2}{1-\frac{2M(t,R)}{R} } 
\nonumber\\
&&\nonumber\\
&{}& +R^2d\Omega_{(2)}^2  \,,\label{generalsphericalmetric}
\end{eqnarray}
where $M(t,R)$ {\em a posteriori} turns out to be the 
Misner-Sharp-Hernandez mass \cite{MisnerSharp, MisnerHernandez}.  
This form is ultimately inspired by  
the  Morris-Thorne wormhole  metric \cite{MorrisThorne},  it  
compromises between the latter  and the widely used gauge 
$ds^2=-A(t,R)dt^2+B(t,R)dR^2 +R^2d\Omega_{(2)}^2$,
and is particularly convenient in the study of both static and 
time-varying black holes \cite{Visser92, NielsenVisser06}.  
For the metric~(\ref{FLRWSchwarzschild}), we have 
\be
\mbox{e}^{-\phi}= \frac{ F(T, R)}{\sqrt{1-kR^2/a^2}} 
\ee
and
\be
1-\frac{2M}{R}=1-\frac{kR^2}{a^2}-H^2R^2 = 
1-\frac{8\pi }{3} \,\rho R^2   
\ee
which is  consistent with the well known expression
\be
M=\left( H^2+\frac{k}{a^2} \right) \frac{R^3}{2} =\frac{4\pi}{3} 
\, R^3 \rho \label{FLRWmass}
\ee 
of the  Misner-Sharp-Hernandez mass 
in FLRW space \cite{Hayward96PRD53}.

In non-spatially flat FLRW spaces, $k\neq 0$, the 
quantity $4\pi R^3/3$ is not the proper  volume of a sphere 
of radius $R$, which is instead
\be
V_{proper}=\int_0^{2\pi}d\varphi \int_0^{\pi}d\theta 
\int_0^{r} dr' \, \sqrt{g^{(3)} } \,,
\ee
where $g^{(3)}= \frac{a^6 r^4 \sin^2\theta}{1-kr^2} $ is the 
determinant of the restriction of the metric $g_{ab}$ to the  
3-surfaces $r=$constant. Therefore, 
\be
V_{proper}=4\pi a^3(t) \int_0^{r} \frac{dr' r'^2}{\sqrt{1-kr'^2}} 
=  
4\pi a^3(t) \int_0^{\chi} d\chi ' f^2(\chi) \,,
\ee
where $\chi$ is the hyperspherical radius and 
\be
f(\chi) =r= \left\{ \begin{array}{ll}
\sinh \chi & \;\;\;\;\;\; \mbox{if} \; k<0 \,,\\
& \\
\chi & \;\;\;\;\;\; \mbox{if} \; k=0 \,, \\
& \\
\sin \chi & \;\;\;\;\;\; \mbox{if} \; k>0 \,, 
\end{array} \right.
\ee
with $ \chi=f^{-1}(r)=\int \frac{dr}{\sqrt{1-kr^2}}$. Integration 
gives
\be
V_{proper}=  \left\{ \begin{array}{ll}
2 \pi a^3(t) \left( r\sqrt{1+r^2} -\sinh^{-1} r \right) & 
\;\;\;\;\;\; \mbox{if} \; k=-1 \,,\\
& \\
\frac{4\pi}{3}\,  a^3(t) r^3 & \;\;\;\;\;\; \mbox{if} \; k=0 \,, 
\\
& \\
2 \pi a^3(t) \left( \sin^{-1} r -r\sqrt{1-r^2} \right) & 
\;\;\;\;\;\; \mbox{if} \; k=+1 \,. 
\end{array} \right.
\ee
However, it turns out that only the ``areal volume'' 
\be\label{arealvolume}
V\equiv \frac{4\pi R^3}{3} 
\ee
is used, as a consequence  of the use of the  
Misner-Sharp-Hernandez mass, which is identified as the internal 
energy $U$ in the thermodynamics of the apparent 
horizon.

The Misner-Sharp-Hernandez mass~(\ref{FLRWmass}) of a sphere of 
radius 
$R$  does not 
depend explicitly on the pressure $P$ of the cosmic fluid.  Its 
time  derivative, instead, depends explicitly on  $P$; consider  
a sphere of proper radius  $R=R_s(t)$, then, using $R\equiv ar$ 
and  eq.~(\ref{FLRWconservation}),  one has
\be
\dot{M}  =  4\pi R_s^3 \left[ \frac{\dot{R}_s}{R_s} \, \rho -H 
\left(  P+\rho \right) \right] \,. 
\ee
If the sphere is comoving, $R_s \propto a(t)$, then 
$\dot{R}_s/R_s=H$ and 
\be
\dot{M}=-4\pi H R_s^3 P \,; \label{FLRWMSHdot}
\ee
in this case $\dot{M}$ depends explicitly on $P$ but not on 
$\rho$. By taking the ratio of  eqs.~(\ref{FLRWMSHdot}) 
and~(\ref{FLRWmass}) one also obtains, in GR,
\be\label{MSHevolution}
\dot{M}+3H \, \frac{P}{\rho} \, M =0  \;\;\;\;\;\;\;\;
(\mbox{comoving sphere}). 
\ee
However, for the thermodynamics of the apparent horizon (and of 
the event horizon as well), the horizon is {\em not} a  comoving 
surface.

It is now easy to locate the apparent horizon of a general FLRW 
space. In spherically symmetric  spacetimes, the apparent horizon 
(existence, location, dynamics, surface gravity, {\em etc.}) can 
be studied by 
using the 
Misner-Sharp-Hernandez mass $M $ 
\cite{MisnerSharp, MisnerHernandez}, which coincides 
with the Hawking-Hayward quasi-local mass 
\cite{Hawking68, Hayward94} for these spacetimes. The 
Misner-Sharp-Hernandez mass is only defined for spherically 
symmetric spacetimes. A spherically 
symmetric line element can always be  written as
\be\label{2normal}
ds^2=h_{ab}dx^a dx^b +R^2 d\Omega_{(2)}^2 \, ,
\ee
where $a,b=1,2$. The 
Misner-Sharp-Hernandez 
mass $M$ is defined by \cite{MisnerSharp, MisnerHernandez}
\be \label{MisnerSharpmass}
 1-\frac{2M}{R} \equiv \nabla^c R  \, \nabla_c R 
\ee
or \footnote{In $(n+1)$ spacetime dimensions, the 
Misner-Sharp-Hernandez mass is $ M=\frac{n(n-1)}{16\pi G} 
R^{n-2} V_n \left( 1-h^{ab}\nabla_a R\nabla_b R \right) $, where 
the line element is $ds^2=h_{ab}dx^a dx^b +R^2 d\Omega^2_{n-1}$ 
($a,b=1,2$) and $V_n= \frac{ \pi^{n/2} }{\Gamma \left( 
\frac{n}{2}+1\right)} $ is the 
volume of the $(n-1)$-dimensional unit ball \cite{BakRey00CQG}.}   
\be M=\frac{R}{2} \left( 1-h^{ab}\nabla_a R \, \nabla_b R 
\right) \,,
\ee
an invariant 
quantity of the 2-space normal to the 2-spheres of symmetry.  
In a FLRW space, setting $g^{RR}=0$ (equivalent to 
$h^{ab}\nabla_a R \, \nabla_b R 
=0$ in Schwarzschild-like coordinates or to $R_{AH}=2M$) yields 
the  radius of 
the FLRW apparent horizon
\be
R_{AH}=\frac{1}{\sqrt{H^2+k/a^2}} \,.
\ee
Since the Misner-Sharp-Hernandez mass is defined quasi-locally 
\cite{Hayward96PRD53}, this derivation illustrates the 
quasi-local nature of 
the apparent horizon, as opposed to the global nature of the 
event and particle horizons. 

Eqs.~(\ref{MisnerSharpmass}) and~(\ref{FLRWSchwarzschild}) yield
\be
1-\frac{2M}{R} =1-H^2R^2- \frac{kR^2}{a^2} 
\ee
and the Hamiltonian constraint~(\ref{Friedmann1}) then implies 
that 
\be
M(R)=\frac{4\pi R^3}{3} \, \rho \,.
\ee

The Kodama vector is introduced as follows. Using the metric 
decomposition~(\ref{2normal}), let $\epsilon_{ab}$ 
be the volume form associated with the 2-metric $h_{ab}$;  then 
the Kodama vector is \cite{Kodama}
\be
K^a \equiv \epsilon^{ab} \nabla_b R 
\ee
with $K^{\theta}=K^{\varphi}=0$.  The Kodama vector lies in the 
2-surface orthogonal to the 2-spheres of symmetry and  
$ K^a \nabla_a R= \epsilon^{ab} \nabla_a R \nabla_b R =0$. In a 
static spacetime, the Kodama vector is parallel (in general, not 
equal) to the timelike Killing vector. In the 
region in which it is timelike, the Kodama vector 
defines a class of preferred observers with four-velocity $ u^a 
\equiv K^a/ \sqrt{ | K^c K_c | } $. It can be proved 
(\cite{Kodama}, see \cite{AbreuVisser10} for a 
simplified proof) that the Kodama vector is 
divergence-free, $ \nabla_a K^a=0 $, which  has the consequence 
that the Kodama energy  current $ J^a \equiv G^{ab}K_b $
is covariantly conserved, $\nabla^aJ_a=0$, a remarkable property 
referred to as the ``Kodama miracle'' \cite{AbreuVisser10}.  If 
the spherically symmetric metric is written in  the gauge
\be\label{ABgauge}
ds^2=-A \left(t,R \right) dt^2+B\left(t,R \right) dR^2+R^2 
d\Omega_{(2)}^2 \,,
\ee
then the Kodama vector assumes the simple form ({\em e.g.}, 
\cite{Racz06})
\be \label{KodamaAB}
K^a=\frac{1}{\sqrt{AB}} \left( \frac{\partial}{\partial 
t} \right)^a \,. 
\ee
It is shown in \cite{Hayward96PRD53} that the Noether charge 
associated with the Kodama current is the  
Misner-Sharp-Hernandez energy \cite{MisnerSharp, MisnerHernandez} 
of spacetime. The Hayward proposal for the horizon surface 
gravity in spherical symmetry  \cite{Hayward98CQG} is based 
on the Kodama vector. This definition is unique because the 
Kodama vector is unique and 
$\kappa_{{\small {\sf Kodama}}}$  agrees with the surface gravity 
on the horizon 
of a Reissner-Nordstr\"om black hole, but not with other 
definitions  of dynamical surface gravity.  The Kodama-Hayward 
surface gravity can be written as  \cite{Hayward98CQG}
\be \label{kappaKodamaHayward}
\kappa_{ {\small {\sf Kodama} } } =\frac{1}{2} \, 
\Box_{(h)}R=\frac{1}{2\sqrt{-h}}  \, \partial_{\mu} 
\left( \sqrt{-h} 
\, h^{\mu\nu} \partial_{\nu}R \right) \,.
\ee
The components of the Kodama vector in Schwarzschild-like 
coordinates are
\be\label{KodamaFLRWSchwarzschild}
K^{\mu}=\left( \frac{ \sqrt{ 1-kR^2/a^2}}{F} , 0,0,0 \right) 
\ee
and its norm squared is 
\be
K_cK^c=-\left( 1-H^2R^2-\frac{kR^2}{a^2} \right)=-\left( 
1-\frac{R^2}{R_{AH}^2 } \right) \,.
\ee
The Kodama vector is timelike ($K_cK^c<0$) if $R<R_{AH}$, null if 
$R=R_{AH}$, and spacelike ($K_cK^c>0$) outside the 
apparent horizon $R>R_{AH}$.  
  
The components of the Kodama 
vector  in  pseudo-Painlev\'{e}-Gullstrand coordinates are
\be
K^{\mu}= \left(  \sqrt{ 1-kR^2/a^2} , 0,0,0 \right) \,,
\ee
while in comoving coordinates they are 
\be
K^{\mu}=\left( \sqrt{1-kr^2}, - Hr\sqrt{1-kr^2}, 0, 0 \right) \,,
\ee
with  $K^c K_c=-\left( 1 -kr^2-\dot{a}^2 r^2 
\right)=1-2M/R  $ ({\em e.g.}, 
\cite{DiCriscienzoHaywardNadaliniVanzoZerbini10}).

In GR, if the  FLRW universe is sourced by a 
perfect fluid with energy-momentum  tensor
\be
T_{ab}=\left( P+\rho \right) u_a u_b +P g_{ab} \,,
\ee
where $\rho$, $P$, and $u^a$ are the energy density, pressure, 
and four-velocity field of the fluid, respectively, one has
\begin{eqnarray}
H^2 & = &   \frac{8\pi G}{3} \, \rho -\frac{k}{a^2} \,, 
\label{Friedmann1} \\
&&\nonumber\\
\frac{ \ddot{a} }{a} & = &   -\, \frac{4\pi 
G}{3}  \left( \rho +3P  \right) \,. \label{Friedmann2} 
\end{eqnarray}
The covariant 
conservation equation $\nabla^b T_{ab}=0$  yields the energy 
conservation equation
\be \label{FLRWconservation}
\dot{\rho}+3H \left( P+\rho \right)=0 
\ee
which  is not independent of   
eqs.~(\ref{Friedmann1}) and (\ref{Friedmann2}) and  
can be derived from them. Another useful relation following from 
these equations is
\be
\dot{H}=-4\pi G \left( P+\rho \right) +\frac{k}{a^2} \,.
\ee

Let $t=0$ denote the Big Bang  singularity (in the cases in 
which it is present).  All comoving observers whose worldlines 
have   $u^a$ as tangent are equivalent and, therefore, the  
following considerations apply to any of them,  although we 
refer explicitly to a comoving observer located at $r=0$.

\section{FLRW horizons and their dynamics}

Two horizons of FLRW space are familiar from standard 
cosmology textbooks:  the particle  and the event 
horizons \cite{Rindler56}.  The {\em particle horizon} 
\cite{Rindler56} at time $t$ is a 
sphere centered 
on the 
comoving observer at $r=0$  and with radius
\be\label{FLRWparticlehorizon}
R_{PH}(t)=a(t) \int_0^t \frac{dt'}{a(t')} \,.
\ee
The particle horizon contains every particle signal that has 
reached the observer between the time of the Big Bang $t=0$ and 
the time $t$ \footnote{More realistically, photons propagate 
freely in the universe only after the time of the  last 
scattering or recombination, before which the Compton 
scattering due to free electrons in the cosmic plasma makes it 
opaque. Therefore, cosmologists introduce  the {\em optical 
horizon} with radius $ a(t) \int_{t_{recombination}}^t 
\frac{dt'}{a(t')}$ \cite{Mukhanovbook}. However, the optical 
horizon is irrelevant for our purposes and will not be used 
here.}.  For particles travelling radially to the 
observer  at light speed, it is $ds=0$ and $d\Omega_{(2)}=0$. The 
line  element can be written using hyperspherical coordinates,
\be\label{FLRWmetric}
ds^2=-dt^2+a^2(t) \left[ d\chi^2 +f^2( \chi ) d\Omega_{(2)}^2 
\right] \,.
\ee
Along  radial  null geodesics, $ d\chi=-dt/a$ and the 
infinitesimal  proper radius is $a(t) d\chi$). Integrating 
between the emission of a light signal at $\chi_e$ at time $t_e$ 
and 
its  detection at $\chi=0$ at time $t$, one obtains 
\be
\int_{\chi_e}^0 d\chi=-  \int_{t_e}^t \frac{dt'}{a(t')} 
\ee
and, using $\chi_e= \int_0^{\chi_e} d\chi=-\int_{\chi_e}^0 d\chi 
$, we obtain \cite{dInvernobook}
\be
\chi_e=\int_{t_e}^t \frac{dt'}{a(t')} \,.
\ee
The physical (proper) radius $R$  is obtained by multiplication 
by the scale factor \footnote{The 
notation  for  the proper radius $R\equiv a(t)r=a(t)f(\chi)$ is  
consistent  with our previous  use of this symbol to denote an 
areal  radius, since $ a(t)r$ is in fact an areal radius, as is 
obvious from the inspection of the FLRW line 
element~(\ref{FLRWmetric}). If $k\neq 0$, the proper radius 
$a(t)\chi$ and the areal radius $a(t) f(\chi)$ do not coincide.},
\be
R_e=a(t) \int_{t_e}^t \frac{dt'}{a(t')} \,,
\ee
Now take the limit $t_e\rightarrow 0^{+}$:
\begin{itemize}
\item if the integral $\int_0^t \frac{dt'}{a(t')} $ diverges, 
it is possible for the observer at $r=0$ to receive all the light 
signals emitted at sufficiently early times from any point in the 
universe. The maximal volume that can be causally connected to 
the observer at time $t$ is infinite.

\item If the integral $\int_0^t \frac{dt'}{a(t')} $ is finite, 
the observer at $r=0$ receives, at time $t$, only the light 
signals started within the sphere $ r \leq \int_0^t 
\frac{dt'}{a(t')} $.

\end{itemize}

The physical (proper) radius of the particle horizon is therefore 
given by eq.~(\ref{FLRWparticlehorizon}). At a given time 
$t$, the particle horizon  is the boundary between the 
worldlines that can be seen by the  observer and those 
(``beyond the horizon'') which  cannot be seen. This boundary 
hides  events which cannot be known by that observer at 
time $t$ and it  evolves with time. The particle horizon is the 
horizon commonly studied in inflationary cosmology.

The particle and event horizons depend on the 
observer: contrary to  the event horizon of the  Schwarzschild 
black hole, different comoving observers in 
FLRW space will see event horizons located at different places.  
Another difference with respect 
to a  black hole horizon is that the observer is located 
{\em inside}  the event horizon and cannot be reached by signals 
sent  from the outside.

The cosmological particle horizon is a null surface.  This 
statement is obvious from the 
fact that the event horizon is a causal boundary and is generated 
by the null geodesics which barely fail to reach the observer; it 
can also be checked explicitly. Using 
hyperspherical coordinates  $\left( t, \chi \right)$, the 
equation of the particle horizon is 
\be
{\cal F}\left( t, \chi \right) \equiv \chi -\int_0^{t} 
\frac{dt'}{a(t')}=0 \,.
\ee
The normal to this surface has components
\be
N_{\mu}  =  \nabla_{\mu} {\cal F} \left. \right|_{PH} = 
\delta_{\mu 1 } -\frac{ \delta_{\mu 0}}{a} 
\ee
and it is straightforward to see that $  N^a N_a = 0$. 

The particle horizon evolves according to the equation 
({\em e.g.}, \cite{DasChattoDebnath11}) 
\be \label{FLRWPHevolution}
\dot{R}_{PH}=HR_{PH}+1 \,,
\ee
which is obtained by differentiating 
eq.~(\ref{FLRWparticlehorizon}).  In an expanding universe with a 
particle horizon it is  
$\dot{R}_{PH}>0$, which means that more and more signals emitted 
between the Big Bang and time $t$ reach the observer as time  
progresses. If $R_{PH}(t)$ does not diverge as $t\rightarrow  
t_{max}$, then there will always be a region unaccessible to 
the comoving observers. 

The acceleration of the particle horizon is
\be
\ddot{R}_{PH} =  \frac{ 
\ddot{a}}{a} \, R_{PH} +H  =   -\frac{4\pi}{3} 
\left( \rho+3P \right) R_{PH} +H \,,
\ee

Let us turn now our attention to the {\em event horizon}. 
Consider all the events which can be seen by the  comoving 
observer at $r=0$ between time $t$ and future infinity 
$t=+\infty$ (in a closed universe which recollapses, or in a Big 
Rip universe which ends at a finite time, substitute $+\infty$ 
with the time $t_{max}$ corresponding to the maximal expansion 
or the Big Rip, respectively). The comoving radius of the region 
which can be seen by this observer is
\be
\chi_{EH}=\int_t^{+\infty} \frac{dt'}{a(t')} \,;
\ee
if this integral diverges as the upper limit of integration goes 
to infinity or to $t_{max}$, it is said that there is no event 
horizon in this FLRW space and events arbitrarily far away can 
eventually be seen by the observer by waiting a sufficiently long 
time. If the integral converges, there is an   event 
horizon: events beyond $r_{EH}$ will never be known to the 
observer \cite{Rindler56}. The physical (proper) radius of the  
event horizon is
\be\label{FLRWeventhorizon}
R_{EH}(t)=a(t)\int_t^{+\infty} \frac{dt'}{a(t')} \,.
\ee
In short, the event horizon can be said to be the 
``complement''of the particle horizon \cite{Mukhanovbook}; it is 
the (proper) distance to the most distant event that the 
observer will ever see. Clearly, in order to define the event 
horizon, one must know the entire future history of the universe 
from time $t$ to infinity and the event horizon is defined {\em 
globally}, not locally.

The cosmological event horizon is a null surface.  Again, the 
statement follows from the  fact that the event horizon is a 
causal boundary. To check  explicitly, use the equation of the 
event horizon in  comoving  coordinates 
\be
{\cal F} \left( t, \chi \right) \equiv \chi 
- \int_t^{t_{max}} 
\frac{dt'}{a(t')}=0 \,;
\ee
the normal to this surface has components
\be
N_{\mu}= \nabla_{\mu} {\cal F} \left. \right|_{EH} = 
\delta_{\mu 1} - \frac{\delta_{\mu 0}}{a} 
\ee
and it is easy to see that $N^a N_a  = 0$.

The event horizon evolves according to the 
equation \cite{AbkarCai06PLB635, MosheniSadjadi06, 
DasChattoDebnath11} 
\be \label{FLRWEHevolution}
\dot{R}_{EH}=HR_{EH} -1 \,,
\ee
which is obtained by differentiating 
eq.~(\ref{FLRWeventhorizon}). The acceleration of the event 
horizon is also straightforward to 
derive,
\be
\ddot{R}_{EH}=\left( \dot{H}+H^2 \right) R_{EH}-H \,.
\ee

The event horizon does not exist in every FLRW 
space. To wit, consider  a spatially flat FLRW universe 
sourced 
by a  perfect fluid with equation of state $P=w\rho$ and 
$w=$const.$> -1$; if  $ w \geq -1/3$ ({\em i.e.}, in GR, 
for a decelerating universe), there is no  event horizon because  
\be
a(t)=a_0\,  t^{  \frac{2}{3(w+1)} } \label{exacta}
\ee
and the event horizon has radius  
\be
R_{EH} =  t^{ \frac{2}{3(w+1)} } \left[ \frac{3(w+1)}{3w+1} \, 
t'^{ \frac{3w+1}{3(w+1)} } \right]_{t}^{+\infty} \,.
\ee
If $w > -1/3 $, the exponent 
$\frac{3w+1}{3(w+1)}$ is positive and the integral diverges: 
there is no 
event horizon in  this case. Indeed, the existence of 
cosmological event horizons 
seems to require the violation of the strong energy condition 
in at least some region of spacetime  
\cite{BhattacharyaLahiri10}. We 
can state that in GR with a perfect fluid the 
event  horizon exists only for accelerated universes with 
$P<-\rho/3$.

The literature sometimes  refers to a ``Hubble horizon'' 
of 
FLRW space with  
radius \be
R_H \equiv \frac{1}{H} \,.
\ee
This quantity only provides the order of magnitude of the 
radius of curvature of a FLRW space and is used as an 
estimate of the radius of the event horizon during 
inflation, when the universe is close to a de Sitter space 
\cite{KolbTurner90}. The Hubble horizon 
coincides with the apparent horizon for spatially flat 
universes (see eq.~(\ref{FLRWapparenthorizon}) below) and with the 
event horizon of de Sitter 
space. However,  this concept does not add  to the discussion 
of the various types of FLRW horizons and it seems unnecessary.

Let us consider now the {\em apparent horizon}, which  
depends on the  spacetime slicing (this feature is illustrated  
by the fact that it is possible to find non-spherical slicings 
of the Schwarzschild spacetime without any apparent horizon 
\cite{WaldIyer91PRD, SchnetterKrishnan06}). In a FLRW spacetime, 
it is  natural to  use a 
slicing with hypersurfaces of homogeneity and isotropy (surfaces 
of constant comoving time). FLRW space is spherically 
symmetric about every point of space and the outgoing and ingoing 
radial null geodesics have tangent fields with comoving 
components
\be 
l^{\mu} = \left( 1,   \frac{\sqrt{1-kr^{2}}}{a(t)}, 0 ,0 \right) 
\,, \;\;\;\;
 n^{\mu} = \left( 1, -\frac{\sqrt{1-kr^{2}}}{a(t)}, 0 ,0 \right) 
  \,, \label{FLRWnullrays}
\ee 
respectively, as is immediately obtained by setting $ p_c p^c=0$ 
for the tangents.   There is freedom to rescale a null vector by 
an 
arbitrary constant (which must be positive if we want to keep 
this vector future-oriented). The choice~(\ref{FLRWnullrays}) 
implies that 
$l^c n_c=-2$.  The more common normalization $l^c n_c=-1$ is 
obtained by dividing both $l^a$ and $n^a$ by  $\sqrt{2}$.

The expansions of the null geodesic congruences are computed 
using the equation
\be \label{expansion} 
\theta_{l} = \left[ g^{ab} + 
\frac{l^ a n^b    +  n^a l^b}{\left( 
-n^{c}l^{d}g_{cd}\right)} \right] \nabla_{a}l_{b} \,.
\ee 
Computing first
\begin{eqnarray}
\nabla_c l^c & = & 3H+\frac{2}{ar}\, \sqrt{1-kr^2} \,, \\
&&\nonumber\\
\nabla_c n^c &= & 3H - \frac{2}{ar}\, \sqrt{1-kr^2} \,, 
\end{eqnarray}
and  using 
\begin{eqnarray*}
&& \sqrt{-g}  =  \frac{a^3 r^2 \sin^2 
\theta}{\sqrt{1-kr^2}} \,, \;\;\;\;\;\;  g_{cd} l^c n^d = -2 \,,\\
&&\nonumber\\
&& \Gamma^c_{00} =  0 \,, \;\;\;\;\;
\Gamma^c_{01}=\Gamma^c_{10}= H\delta^{ c1} \,, \;\;\;\;\;
\Gamma^c_{11}= \frac{kr \delta^{c1} +a\dot{a} \, \delta^{c 0}}{ 
1-kr^2 } \,,
\end{eqnarray*}
the result is \footnote{The factor~2 in 
eqs.~(\ref{FLRWexpansionout}) 
and~(\ref{FLRWexpansionin}) does not appear in 
Ref.~\cite{BakRey00CQG} because of a different normalization of 
$l^a$ and $n^a$.}
\begin{eqnarray}
\theta_{l} & = &  
\frac{ 2\left( \dot{a}r + \sqrt{1-kr^2 }\right) }{ar} 
=2\left( H + \frac{1}{R}\, \sqrt{ 1-\frac{kR^2}{a^2}} \right)  
\,,\nonumber\\
&& \label{FLRWexpansionout}\\
\theta_{n} & = &  
\frac{2 \left( \dot{a}r - \sqrt{1-kr^2 }\right)}{ar}  
=2\left( H - \frac{1}{R}\, \sqrt{1-\frac{kR^2}{a^2}} \right) 
\,.\nonumber\\
&& \label{FLRWexpansionin}
\end{eqnarray} 
Following \cite{Hayward:1993wb}, the apparent horizon  is a  
surface defined by the  conditions on the time slicings  
\begin{eqnarray} 
& \theta_{l} > 0 \,,& \label{AHcondition1} \\
& \theta_{n} = 0 \,,&  \label{AHcondition2} 
\end{eqnarray}
and is located at  
\be 
r_{AH} = \frac{1}{ \sqrt{\dot{a}^2 + k}} 
  \,, 
\ee
or
\be\label{FLRWapparenthorizon}
 R_{AH}(t)= \frac{1}{\sqrt{H^2+ k/a^2}} 
\ee 
in terms of the proper radius $R\equiv ar$. The 
apparent horizon is defined locally using null geodesic 
congruences and their expansions, and there is no reference to 
the global causal structure. 

Looking at eqs.~(\ref{FLRWexpansionout}) 
and~(\ref{FLRWexpansionin}), or at their product
\be
\theta_l\, \theta_n = \frac{4}{R^2} \left( \frac{R^2}{R_{AH}^2} 
-1 \right) \,,
\ee
it is clear that  when 
$R>R_{AH}$ it is $\theta_l>0$ and $\theta_n>0$, while the region 
$0 \leq R <R_{AH} $ has $ \theta_l>0$ and $\theta_n  <0$ 
(radial null rays coming from the region outside the horizon 
will not cross it and reach the observer).

For a spatially flat universe, the radius of the  
apparent horizon $R_{AH}$ coincides with the Hubble radius 
$H^{-1}$, while  for a positively curved ($k>0$) universe 
$R_{AH}$  is smaller than the  Hubble radius, and it is larger 
for an open ($k<0$) universe. In GR, the Hamiltonian  constraint~(\ref{Friedmann1}) 
guarantees that the  argument of the square root in 
eq.~(\ref{FLRWapparenthorizon}) 
is positive for positive densities $\rho$. The apparent horizon 
exists in all FLRW spaces.

In general, the apparent horizon is not a null 
surface, contrary to the event and particle horizons.    
The equation of the apparent horizon in comoving 
coordinates is 
\be
{\cal F}(t,r)= a(t) \, r -\frac{1}{\sqrt{ H^2+k/a^2} } =0 \,.
\ee
The normal has components
\begin{eqnarray}
N_{\mu} & = & \nabla_{\mu} {\cal F} \left. \right|_{AH} 
\nonumber\\
&&\nonumber\\
&=&  
 \left\{ \left[ \dot{a}r +\frac{ H \left( \dot{H}-k/a^2 
\right) }{ \left( H^2+k/a^2 \right)^{3/2}} \right] \delta_{\mu 
0}+a  \, \delta_{\mu 1} \right\}_{AH} \nonumber\\
&&\nonumber\\
&=& HR_{AH} \left[ 1+\left( \dot{H}-\frac{k}{a^2} \right) 
R_{AH}^2 \right] \delta_{\mu 0}+a \delta_{\mu 1} \nonumber\\
&=& HR_{AH}^3 \, \frac{\ddot{a}}{a} \, \delta_{\mu 0} +a \, 
\delta_{\mu 1} \,.  
\end{eqnarray}
In GR with a perfect fluid with equation of 
state $P=w\rho $, eqs.~(\ref{Friedmann1}), (\ref{Friedmann2}), 
and ({\ref{FLRWapparenthorizon}) yield
\be\label{4point29}
N_{\mu}=-\frac{ \left( 3w+1\right)}{2} \, HR_{AH} \delta_{\mu 
0}+a\delta_{\mu 1} \,.
\ee
The norm squared of the normal is 
\begin{eqnarray}
N^a N_a &=&   1 - kr^2_{AH} - \frac{ H^2 \left( \dot{H}+H^2 
\right)^2 }{\left( H^2+k/a^2\right)^3 } \nonumber\\
&&\nonumber\\
&=&  H^2R_{AH}^2 \left[ 1- \left(\frac{\ddot{a}}{a}\right)^2 
 R_{AH}^4 \right]    \nonumber\\
&&\nonumber\\
&=& \frac{3H^2R_{AH}^2}{4\rho^2} \left( \rho + P\right) 
\left(\rho-3P \right) \\
&&\nonumber\\
&=& H^2R_{AH}^2 \left( 1-q^2 H^4 R_{AH}^4 \right)  \,,
\end{eqnarray}
where $q\equiv -\ddot{a}a/\dot{a}^2$ is the deceleration 
parameter. The horizon is null if and only if $P=-\rho$ or 
$P=\rho/3$. In 
GR with a perfect fluid, the 
Hamiltonian  constraint~(\ref{Friedmann1}) yields 
\be \label{callit44}
H^2R_{AH}^2= \left( 1+\frac{k}{a^2H^2} \right)^{-1} = 
\left( \frac{8\pi G}{3H^2} \, \rho \right)^{-1} 
\equiv 
\frac{\rho_c}{\rho} \equiv \Omega^{-1}\,,
\ee
where $\rho_c\equiv \frac{ 3H^2}{8\pi G}$ is the critical density 
and $\Omega \equiv \rho/\rho_c$ is the density parameter, and one 
obtains
\be \label{4point36}
N^a N_a =  -\frac{3}{4} \left( w+1 \right) 
\left( 3w-1 \right) H^2R_{AH}^2 
=  \frac{ \Omega^2-q^2}{ \Omega^3 } \,.
\ee
Eq.~(\ref{4point36}) establishes that:

\begin{itemize}
\item if $ -1<w<1/3$,  then  
$N^cN_c>0$ and  the apparent 
horizon is {\em timelike}. For a $k=0$ universe in Einstein's 
theory  this  condition corresponds to $ \dot{H}<0$.

\item If $ w=-1 $ or $w=1/3$, then 
$N^c N_c=0$ and the apparent horizon 
is {\em null} (de Sitter space, which has $\dot{H}=0$ and 
$q=-1$, falls into this category but it is not the only space 
with these properties).

\item If  $ w<-1 $ or $w>1/3$, then 
$N^cN_c<0$,  the normal is timelike, and the 
apparent horizon is {\em spacelike}. In Einstein's theory with 
$k=0$ and a perfect fluid as the source,  $w<-1$   corresponds to 
$\dot{H}>0$ (``superacceleration'').  This is the case of Big 
Rip universes and of a phantom fluid  which violates 
the weak energy condition.
\end{itemize}

The black hole dynamical horizons considered in the literature 
are usually required to be {\em spacelike} 
\cite{AshtekarKrishnan}. However, cosmological horizons can be 
timelike. In GR, the radius of the apparent 
horizon can be  written as $ R_{AH}= \left( \sqrt{\Omega} \,  
\left| H \right| \right)^{-1} $  in terms  of the density 
parameter $\Omega$ by using eq.~(\ref{callit44}).

The apparent horizon evolves according to the 
equation \cite{CaiKim05JHEP, AkbarCai07PRD75, 
ChakrabortyMazumderBiswas10, MazumderBiswasChakraborty11, 
DasChattoDebnath11} 
\be\label{AHevolution}
\dot{R}_{AH}=HR_{AH}^3\left( \frac{k}{a^2}-\dot{H} 
\right) = 4\pi HR_{AH}^3 \left( P+\rho \right)\,,
\ee
as is easy to check by  differentiating 
eq.~(\ref{FLRWapparenthorizon}) with 
respect to $t$. In GR with a perfect fluid as a source, the only 
way 
to obtain a stationary apparent horizon is when $P=-\rho$. 
For de Sitter space, eq.~(\ref{AHevolution}) reduces to $ 
\dot{R}_{AH}=0$,  consistent with $R_{H}=H^{-1}$ and $H=$const.  
For non-spatially flat universes, the equation of state 
$P=-\rho$ produces other solutions. For example, for $k=-1$ and a 
cosmological constant $\Lambda>0$ as the only source of gravity, the 
scale factor 
\be
a(t)=\sqrt{\frac{3}{\Lambda}} \, \sinh \left( \sqrt{\frac{\Lambda}{3}} 
\, t \right)
\ee
is a solution of the Einstein-Friedmann equations. The 
radius of the event horizon has the time dependence
\be
R_{EH}(t)=\sqrt{ \frac{3}{\Lambda}} \, \sinh \left( 
\sqrt{\frac{\Lambda}{3}} \, t \right) \left| \ln \left[ \tanh \left( 
\sqrt{\frac{\Lambda}{3}} \, \frac{t}{2} \right) \right] \right| 
\,.
\ee
The apparent horizon, instead, has constant radius 
$R_{AH}=\sqrt{3/\Lambda}$.

As another example consider, for $k=+1$ and cosmological constant 
$\Lambda>0$, the scale factor
\be
a(t)= \sqrt{\frac{3}{\Lambda}} \, \cosh\left( \sqrt{\frac{\Lambda}{3} 
}\, t \right) \,;
\ee
the event horizon has radius
\begin{eqnarray}
R_{EH}(t)& = &  \sqrt{\frac{3}{\Lambda}} \, \cosh \left( 
\sqrt{ \frac{\Lambda}{3}} \, t \right) \nonumber\\
&&\nonumber\\
& \cdot & \left[ 
\frac{\pi}{2} +n\pi 
-\tan^{-1} \left( \sinh \left( \sqrt{\frac{\Lambda}{3}} \, t \right)  
\right) \right] 
\end{eqnarray}
where $n=0, \pm 1, \pm 2 , \; ...$ The multiple possible values of $n$ 
correspond to the infinite possible branches which one can consider 
when inverting the tangent function, and to the fact that in a closed 
universe light rays can travel multiple times around the universe. In 
this situation it is problematic to regard the event horizon as a true  
horizon  \cite{Davies88CQG}. The apparent horizon has constant radius
$ R_{AH}=\sqrt{3/\Lambda}$, according to the fact that 
$\rho_{\Lambda}+P_{\Lambda}=0$ in 
eq.~(\ref{AHevolution}) \footnote{These two examples, together 
with de  Sitter space for $k=0$,  are presented in 
Ref.~\cite{Davies88CQG}.  However, contrary to what is stated in 
this reference, in both cases the  event horizon is not constant: 
it is the apparent horizon instead which is constant.}.

In a $k=0$ FLRW universe with a perfect fluid and constant 
equation of state $P=w\rho$ and $ -1<w<-1/3$  (accelerating but 
not superaccelerating universe), the  event horizon is always 
outside the apparent horizon and is, 
therefore, unobservable \cite{Bousso05PRD, WangGongAbdalla06}.

Let us summarize the dynamical evolution of the 
FLRW horizons  and compare their evolutionary laws. The first 
question to ask is whether these 
horizons are comoving: they almost never are. The difference 
between the expansion rate  of a horizon $ \dot{R}/R$ 
and that of the expanding matter $H$ is, for the particle, 
event, and apparent horizons
\begin{eqnarray}
 \frac{ \dot{R}_{PH} }{ R_{PH} } -H & = &  \frac{1}{R_{PH}} \,,\\
&&\nonumber\\
\frac{ \dot{R}_{EH} }{R_{EH}} -H & = &  -\,\frac{1}{R_{EH}} 
\,,\\
&&\nonumber\\
\frac{ \dot{R}_{AH} }{R_{AH}} -H & = &  H\left[  \frac{ \left( 
\frac{k}{a^2} -\dot{H} \right)}{H^2+\frac{k}{a^2} } -1\right]  =
-\left(  \frac{ \ddot{a}}{a} \, H \right)R_{AH}^2  \nonumber\\
&&\nonumber\\
& = &  \frac{\left(3 w+1 \right)H}{2} \,,\
\end{eqnarray}
respectively.  Taking into consideration only expanding FLRW 
universes ($H>0$), when it exists 
the  particle horizon  always expands faster than comoving. The 
event horizon (which only exists for accelerated 
universes) always expands slower than comoving. The 
apparent horizon expands faster than comoving for decelerated 
universes ($\ddot{a}<0$); slower than comoving for 
accelerated universes ($\dot{a}>0$); and comoving for 
coasting universes ($a(t) \propto t$).

An even simpler  way of looking at the evolution is by using the 
comoving  radius of the horizon: if this radius is constant, 
then the horizon is comoving. We have,
\begin{eqnarray}
&& \dot{r}_{PH}=\frac{1}{a} >0 \,,\\
&&\nonumber\\
&& \dot{r}_{EH}=-\frac{1}{a}  <0 \,,\\
&&\nonumber\\
&& \dot{r}_{AH}= - \frac{\dot{a} \ddot{a} }{ \left( \dot{a}^2+k  
\right)^{3/2}  }  \,,
\end{eqnarray}
respectively.  
The causal character and the dynamics of the various FLRW 
horizons are summarized in 
Tables~\ref{Tablehorizoncharacter} 
and~\ref{Tablehorizondynamics}.

\begin{table}[h]
\rotatebox{90}{\vbox to \textwidth{
\hsize=\textheight
\begin{tabular}{||c|c||}
\hline
\hline
&\\
Horizon &  causal character \\
\hline
\hline
&\\
Event horizon  & null  \\ 
&\\
\hline
&\\
Particle horizon & null \\ 
&\\
\hline
&\\
Apparent horizon  & $\begin{array}{ll} 
\mbox{timelike if} \;\; -\rho<P<\rho/3,\\ 
\mbox{null if} \;\; P=-\rho \; \mbox{or} \; \rho/3,\\
\mbox{spacelike if} \;\; P<-\rho \;\; \mbox{or}\,\, 
P>\rho/3
\end{array} $ \\ 
&\\
\hline
&\\
de Sitter horizon & null   \\  
&\\
\hline
\hline
	\end{tabular} 
\caption{Causal character of the FLRW cosmological horizons.}
\label{Tablehorizoncharacter}
}}
\end{table}

\begin{table}[p]
\rotatebox{90}{\vbox to \textwidth{
\hsize=\textheight
\begin{tabular}{||c|c|c|c||}
\hline
\hline
&&&\\
Horizon & location & velocity &  acceleration
\\
\hline
\hline
&&&\\
Event horizon & $R_{EH}=a(t)\int_t^{+\infty} 
\frac{dt'}{a(t')} $  & $ \dot{R}_{EH}=HR_{EH}-1$  & 
$\ddot{R}_{EH}=\frac{\ddot{a}}{a}R_{EH}-H $ \\ 
&&&\\
\hline
&&&\\
Particle horizon & $R_{PH}= a(t)\int_0^t  
\frac{dt'}{a(t')} $  & $ \dot{R}_{PH}=HR_{PH}+1$  &  
$\ddot{R}_{PH}=\frac{\ddot{a}}{a}R_{PH}+H $ \\ 
&&&\\
\hline
&&&\\
Apparent horizon & $ R_{AH}= \frac{1}{\sqrt{ 
H^2+k/a^2}}  $  & $\begin{array}{ll} 
\dot{R}_{AH} =  H R_{AH}^3 \left(  \frac{k}{a^2}-1 \right)\\
\\
\;\;\;\;\;\; = 4\pi HR_{AH}^2 (P+\rho)  
\end{array}$ 
& $ \begin{array}{ll}
\ddot{R}_{AH}=R_{AH}^3 \\
\cdot \left[ \frac{k}{a^2}\left( 
\dot{H}-H^2\right) \right. \\
\left. -\dot{H} +3H^2 R_{AH}^2 \right.\\
\left. \cdot \left( 
\frac{k}{a^2}-1\right)\right] 
\end{array}$ \\ 
&&&\\
\hline
&&&\\
de Sitter horizon & $R_{dS}=H^{-1} $  & $ \dot{R}_{dS}=0 $  
& $\ddot{R}_{dS}=0$ \\  
&&&\\
\hline
\hline
	\end{tabular} 
\caption{FLRW cosmological horizons and their dynamical behaviour.}
\label{Tablehorizondynamics}
}}
\end{table}

\section{Trapping horizon of FLRW space}

Let us now ask the question:  When is the  FLRW apparent horizon 
also a trapping horizon? 
According to Hayward's definition, when 
${\cal L}_l \theta_n>0$, which gives the coordinate-  
(but not slicing)-invariant criterion  
\be \label{FLRWTH}
{\cal L}_l \theta_n=\frac{{R^a}_a}{3} >0 \,,
\ee
where ${R^a}_a$ is the Ricci scalar of FLRW space, and ${\cal 
L}_l$ is the Lie derivative along $l^a$. In fact, using 
eqs.~(\ref{FLRWnullrays}) and~(\ref{FLRWexpansionin}), we have  
\begin{eqnarray*}
{\cal L}_l \theta_n & = & l^a\nabla_a \theta_n= l^a\partial_a 
\theta_n \\
&&\\
&=& 2 \left( \partial_t +\frac{ \sqrt{1-kr^2}}{a} \, \partial_r 
\right) \left( H-\frac{ \sqrt{1-kr^2} }{ar} \right)\\
&&\\
&=& \frac{2}{R^2} \left( \dot{H}R^2 + HR\sqrt{1-\frac{kR^2}{a^2} }  
+ 1 \right) \,.
\end{eqnarray*}
At the apparent horizon $R=R_{AH}$ it is
\begin{eqnarray*} 
&& {\cal L}_l \theta_n \left. \right|_{AH}  \\
&&\\
&& = \, 
2\left( H^2+\frac{k}{a^2} \right)\left[ 
\frac{\dot{H}}{H^2+\frac{k}{a^2} }  + \frac{ H\sqrt{ 
1-\frac{k}{a^2(H^2+k/a^2)} }} {\sqrt{ H^2+k/a^2}} +1 \right] \\
&&\\
&& =2\left( \dot{H}+2H^2+\frac{k}{a^2} \right) 
=\frac{{R^a}_a}{3} \,.
\end{eqnarray*}
This result is independent of the field equations. If we assume 
Einstein's theory and a perfect fluid as the sole source of 
gravity, we obtain 
\be
{\cal L}_{l} \theta_n \left. \right|_{AH} =  
\, \frac{8\pi G}{3} \left(  \rho-3P \right)
\ee
and, therefore,

\medskip
\medskip
\noindent {\em the apparent horizon is also a trapping 
horizon iff ${R^a}_a>0$  (equivalent to $P<\rho/3$ in GR  with a 
perfect fluid).} 

\medskip

Note that in a   radiation-dominated universe, which is 
decelerated, the event  horizon does not exist.

\section{Thermodynamics of cosmological horizons in GR}

Originally developed for static or stationary {\em event} 
horizons, black hole thermodynamics has now been extended to  
apparent, trapping, isolated, 
dynamical, and  slowly evolving  horizons \cite{Boothreview, 
Nielsenreview, AshtekarKrishnan}. 
Similarly, the cosmological thermodynamics associated with 
cosmological horizons has been extended from the 
static  de Sitter event horizon \cite{GibbonsHawking77}  
to FLRW dynamical {\em apparent} horizons.

The thermodynamic formulae valid for the de Sitter event (and 
apparent) horizon are generalized to the non-static apparent 
horizon of FLRW 
space.  The apparent 
horizon is argued to 
be a causal horizon associated with gravitational temperature, 
entropy and surface gravity in dynamical spacetimes 
(\cite{Collins92, HaywardMukohyamaAshworth99PLA, Hayward98CQG, 
BakRey00CQG, Bousso05PRD, NielsenYeom09} and references therein) 
and these arguments apply also to cosmological horizons. 
That thermodynamics are ill-defined for the event horizon of 
FLRW space was argued in \cite{Davies88CQG, FrolovKofman03, 
Bousso05PRD, 
WangGongAbdalla06}. The Hawking 
radiation of the FLRW apparent horizon was computed in 
\cite{ZhuRen08, JangFengPeng09}. The authors of 
\cite{Medved02, CaiCaoHu09} rederived it using the 
Hamilton-Jacobi method  \cite{Visser03IJMPD, NielsenVisser06, 
AnghebenNadaliniVanzoZerbini05}  in the Parikh-Wilczek 
approach originally developed for black 
hole horizons \cite{ParikhWilczek00}. In this context, the 
particle emission rate in the WKB approximation is the tunneling 
probability for the classically forbidden trajectories from 
inside to outside the horizon,
\be
\Gamma \sim \exp\left( -\frac{2 \, \mbox{Im}(I)}{\hbar} \right)  
\simeq 
\exp \left(  - \, \frac{\hbar \omega}{K_B T}  \right) \,,
\ee
where $ I $ is the Euclideanized action with imaginary part 
$\mbox{Im}( I ) $, $\omega$ is the angular frequency of the 
radiated quanta 
(taken, for simplicity, to be those of a massless scalar field, 
which is the simplest field to perform Hawking effect 
calculations), and the Hawking 
temperature is read off the expression of the Boltzmann factor,
$ K_BT=\frac{\hbar \omega}{2 \, \mbox{Im}( I )}$. The particle 
energy $\hbar 
\omega$ is defined in an invariant way as $\omega =-K^a\nabla_a 
I$, where $K^a$ is the Kodama vector, and the action $I$ 
satisfies the Hamilton-Jacobi equation
\be
h^{ab}\nabla_a I \, \nabla_b I =0 \,.
\ee
Although the definition of energy is coordinate-invariant, it 
depends on the choice of time, here defined as the Kodama time.

 A review of the thermodynamical properties 
of the FLRW apparent horizon, as well as the computation of the 
Kodama vector, Kodama-Hayward surface gravity, and Hawking 
temperature in various coordinate systems are given in 
Ref.~\cite{DiCriscienzoHaywardNadaliniVanzoZerbini10}.  The  
Kodama-Hayward 
temperature of the FLRW apparent horizon is given by
\begin{eqnarray}  
K_B T & = & \left( 
\frac{\hbar}{c} \right) \frac{ R_{AH} \left( 
H^2+\frac{\dot{H}}{2}+\frac{k}{2a^2} \right)}{2\pi} \nonumber\\
&&\nonumber\\
&=&  \left( \frac{\hbar}{24\pi c} \right) R_{AH} {R^a}_a 
=  \left( \frac{\hbar G}{c} \right) \frac{R_{AH}}{3} 
\left( 
\rho-3P \right) \,.\nonumber\\
&& \label{KodamaHaywardT}
\end{eqnarray}
The expression of the temperature depends on the choice of 
surface gravity $\kappa$ since $T=|\kappa|/2\pi$ in geometrized 
units and there are several inequivalent 
prescriptions for this quantity (see \cite{Nielsen:2007ac, 
PielahnKunstatterNielsen11} for reviews). The 
choice of $\kappa$ giving 
the temperature reported here is the  Kodama-Hayward  
prescription~(\ref{kappaKodamaHayward})  
\cite{DiCriscienzoHaywardNadaliniVanzoZerbini10}. In fact, this 
equation yields 
\be\label{kappakodama}
\kappa_{ {\small {\sf Kodama}}}=-\frac{R_{AH}}{2} \left( 
2H^2+\dot{H}+\frac{k}{a^2} \right)=-\frac{R_{AH}}{2} \, {R^a}_a 
\,,
\ee
as can be quickly assessed by using comoving coordinates and the 
decomposition~(\ref{2normal}) of the metric, where 
$h_{ab}=$diag$\left( -1, \frac{a^2}{1-kr^2} \right)$.
 The entropy of the FLRW apparent horizon is
\be
S_{AH} = \left( \frac{K_B c^3}{\hbar G} \right) 
\frac{A_{AH}}{4}= 
\left(\frac{K_B c^3}{\hbar G} \right) \frac{\pi}{H^2+k/a^2} \,,
\ee
where   
\be
A_{AH}=4\pi R_{AH}^2 = \frac{4\pi}{H^2 +k/a^2}  
\ee
is the area of the event horizon. The Hamiltonian 
constraint~(\ref{Friedmann1}) gives 
\be
 S_{AH} =  \frac{3}{8\rho} \,, \;\;\;\;\;\;
\dot{S}_{AH} =  \frac{9H}{8\rho^2}\left(P+\rho 
\right)  \,.
\ee
 In an expanding universe the apparent horizon 
entropy increases if $P+\rho>0$, stays constant if $P=-\rho$, 
and decreases if the weak energy condition is violated, 
$P<-\rho$. 

It seems to have gone unnoticed in the literature that the 
horizon temperature  is positive if and only if the Ricci scalar 
is, which is equivalent to  equations of state satisfying  $P<\rho/3$ 
for 
a perfect fluid in Einstein's theory. This is the condition for 
the apparent horizon to be also a trapping horizon. A ``cold 
horizon'' with $T=0$ is obtained for vanishing Ricci scalar but 
the entropy is positive for such an horizon, a situation 
analogous to that of extremal black hole horizons in GR. Note 
also that, if the weak energy condition (which implies $P+\rho 
\geq 0$) is assumed, the boundary $P=\rho/3$ between positive and 
negative Kodama-Hayward temperatures corresponds to the boundary 
between timelike and spacelike character of the apparent 
horizon. It is not obvious {\em a priori} that a null apparent 
horizon, obtained for ${R^a}_a=0$ ($P=\rho/3$ for a perfect 
fluid in GR), should occur when the universe is 
filled with conformal matter.

The natural choice of surface 
gravity seems to be that of Kodama-Hayward, which produces the 
apparent horizon temperature~(\ref{KodamaHaywardT}). 
The apparent horizon entropy is $S_{AH}=\frac{A_{AH}}{4}=\pi 
R_{AH}^2$ (this can be obtained using Wald's Noether charge 
method---see the discussion of the next section), and the 
internal energy $U$ should be identified with 
the Misner-Sharp-Hernandez mass $M_{AH}=\frac{4\pi R_{AH}^3}{3} 
\, \rho$ contained inside the apparent horizon.  The factor 
$4\pi R_{AH}^3/3$  is not the proper 
volume of a sphere of proper (areal) radius $R_{AH}$ unless the 
universe has flat spatial sections. It is the use of the 
Misner-Sharp-Hernandez mass which points us to 
use the areal volume $
V_{AH} \equiv \frac{4\pi R_{AH}^3}{3} $  instead of the proper 
volume  when discussing thermodynamics (failing to do so would 
jeopardize 
the possibility of writing the 1st law consistently). However, 
even with this 
{\em caveat}, the 1st law does  not assume the form 
\be\label{wrongfirstlaw}
T_{AH}\dot{S}_{AH}=\dot{M}_{AH}+P\dot{V}_{AH}
\ee
that one might expect.  Let us review now the laws of 
thermodynamics for cosmological horizons.

\medskip 
\noindent {\em 0th law.}~The temperature (or, equivalently, the 
surface gravity) is constant on the horizon. This law ensures 
that all points of the  horizon are at the same temperature, or 
that there is no temperature gradient on it. The 0th law is a 
rather trivial consequence of spherical symmetry.

\medskip 
\noindent {\em 1st law.}~ 
The 1st law of thermodynamics for apparent horizons is more 
complicated than~(\ref{wrongfirstlaw}) and was given in 
Refs.~\cite{Hayward98CQG, HaywardMukohyamaAshworth99PLA} under 
the name of  ``unified 1st law''. While using the 
Misner-Sharp-Hernandez mass $M_{AH}$ as internal energy, the 
Kodama-Hayward horizon temperature~(\ref{KodamaHaywardT}), and 
the 
areal volume, one introduces further 
quantities as follows \cite{Hayward98CQG, 
HaywardMukohyamaAshworth99PLA}. Decompose the metric as in  
eq.~(\ref{2normal}); then the {\em work density} is
\be\label{workdensity}
w \equiv -\frac{1}{2} \, T_{ab}h^{ab} \,;
\ee
\be\label{energysupplyvector}
\psi_a \equiv {T_a}^b \nabla_b R+w\nabla_a R
\ee
is the energy flux across the apparent 
horizon, when computed on this hypersurface. The quantity $A_{AH} 
\psi_a$ is called the {\em energy supply vector}. The quantity
\be
j_a \equiv \psi_a +wK_a
\ee
is a divergence-free energy-momentum vector which can be used in 
lieu of $\psi_a$. The Einstein equations then give
\cite{Hayward98CQG, HaywardMukohyamaAshworth99PLA} 
\begin{eqnarray}
&& M  =  \kappa R^2 +4\pi R^3 w \,,\\
&&\nonumber\\
&& \nabla_a M = A j_a \,.
\end{eqnarray}
The last equation is rewritten as 
\cite{Hayward98CQG, HaywardMukohyamaAshworth99PLA}
\be
A\psi_a=\nabla_a M -w \nabla_a V_{AH}
\ee
(``unified 1st law''). The energy supply vector is then written 
as 
\be
A\psi_a =\frac{\kappa}{2\pi} \nabla_a \left( \frac{A}{4} \right) 
+ R \, \nabla_a \left( \frac{M}{R} \right) \,.
\ee
Along the apparent horizon,  it is $
M_{AH}=R_{AH} /2 $  and 
\be
A_{AH}\psi_a=\frac{\kappa}{2\pi} \nabla_a S_{AH}= T_{AH}\nabla_a 
S_{AH} \,.
\ee
This equation is interpreted by saying that the energy supply 
across the apparent horizon $A_{AH}\psi_a$ is the ``heat'' 
$T_{AH}\nabla_a S_{AH}$ gained. Writing the energy supply 
explicitly gives
\be\label{starn}
T_{AH} \nabla_a S_{AH}=\nabla_a M_{AH} -w\nabla_a V_{AH} 
\ee
and $-w\nabla_aV_{AH}$ is a work term. The ``heat'' entering the 
apparent horizon goes into changing the internal energy $M_{AH}$ 
and performing work due to the change in size of this 
horizon.

Let us compute now the time component of eq.~(\ref{starn}) in 
comoving coordinates for a FLRW space sourced by a perfect fluid 
in GR. We have 
\be
w \equiv -\frac{1}{2} \left[ \left( P+\rho \right) 
u_au_b+Pg_{ab} \right] h^{ab} = \frac{\rho -P}{2} \,,
\ee
\be
\dot{V}_{AH} =  3HV_{AH}\left(1-\frac{\ddot{a}}{a} \, R_{AH}^2 \right) 
= \frac{9HV_{AH}}{2\rho} \left(P+\rho \right) \,,
\ee
\be
\dot{M}_{AH} = \frac{d}{dt} \left( V_{AH}\rho \right) 
=  \frac{3HV_{AH}}{2}  \left( P+\rho  \right) \,, 
\ee
and 
\be\label{SAHdot}
\dot{S}_{AH}=2\pi R_{AH}\dot{R}_{AH}=  \frac{3\pi 
R_{AH}^2}{\rho} \, H\left(P+\rho \right) 
\ee
so that 
\begin{eqnarray}
T_{AH}\dot{S}_{AH} &= & \frac{HV_{AH}}{2}\left( 1-\frac{ 
\ddot{a}}{a} \, 
R_{AH}^2 \right)\left( \rho-3P \right) \nonumber\\
&&\nonumber\\
&=& \frac{3HV_{AH}}{4\rho} \left( P+\rho \right)\left( \rho-3P 
\right) \,.
\end{eqnarray}
Therefore, it is  \cite{Hayward98CQG, 
HaywardMukohyamaAshworth99PLA} 
\be\label{1stlaw}
T_{AH}\dot{S}_{AH} = \dot{M}_{AH}+\frac{ \left( P-\rho 
\right)}{2} \, \dot{V}_{AH} \,. 
\ee
In the infinitesimal interval of comoving time $dt$ the changes 
in the thermodynamical quantities are related by 
\be
T_{AH}dS_{AH}=dM_{AH}+dW_{AH} \,, \;\;\;\;\;
dW_{AH}=\frac{\left(P-\rho \right)}{2} \, dV_{AH} \,.
\ee
The coefficient of $dV_{AH}$, {\em i.e.}, $-w=\left(P-\rho 
\right)/2$ equals the pressure $P$ (the naively expected 
coefficient) only if $P=-\rho$ (which includes de Sitter space 
in which $dM_{AH}, dV_{AH}$, and $dS_{AH}$ all vanish). The fact 
that the coefficient appearing in the work term is not simply $P$ 
can be understood as a consequence of the fact that the apparent 
horizon is not comoving. For a comoving sphere of radius $R_s$ it 
is $\dot{R}_s/R_s=H$ and $\dot{V}_s=3HV_s$, while 
\be
\dot{M}_s=\dot{V}_s\rho+V_s\dot{\rho}=3HV_s\rho-3HV_s\left(P+\rho 
\right) =-3HV_s P \,,
\ee
hence $\dot{M}_s+P\dot{V_s}=0$. Indeed, the covariant 
conservation equation~(\ref{FLRWconservation}) is often presented 
as the 1st law of thermodynamics for a comoving volume $V$. 
Because of spatial homogeneity and isotropy there can be no 
preferred directions and physical spatial vectors in FLRW space, 
therefore the heat flux through a comoving volume must be zero. 
In fact, consider a comoving volume $V_c$ (which, by definition,   
is constant in time) and the corresponding proper 
volume at time $t$, $V=a^3(t) V_c$. Multiplying 
eq.~(\ref{FLRWconservation}) by $V$ one obtains
\be
V \dot{\rho}+ \dot{V} \left( P+\rho \right)=\frac{d}{dt} \left( 
\rho V \right)+P\dot{V} =0 \,.
\ee
By interpreting $U \equiv \rho V$ as the total internal energy of 
matter in $V$, one obtains the relation between 
variations in the time $dt$
\be
dU+PdV=0 \,,
\ee
and the 1st law  (with work term coefficient $P$) then gives 
$TdS=0$, which is consistent with the above-mentioned absence of 
entropy flux vectors and with the well known fact that, in curved 
space, there is  no entropy generation in a perfect fluid (the 
entropy along fluid  lines remains constant and there is no 
exchange of entropy  between neighbouring fluid lines 
\cite{Stephani}). Indeed, eq.~(\ref{MSHevolution}) for the 
evolution of the 
Misner-Sharp-Hernandez mass contained in a {\em comoving}  sphere 
reduces to $\dot{M}+P\dot{V}=0$ or $\dot{\rho}+3H\left(P+\rho 
\right)=0$.  However, for a non-comoving volume, the work 
term is more complicated than $PdV$.

Attempts to write the 1st law for the event, instead of the 
apparent, horizon lead to inconsistencies 
\cite{Davies88CQG, FrolovKofman03, 
Bousso05PRD, WangGongAbdalla06}. This fact supports the belief 
that it is the apparent horizon which is the relevant quantity in 
the thermodynamics of cosmological horizons.

\medskip 

\noindent {\em (Generalized) 2nd law.}~A second 
law of thermodynamics for the event horizon of de 
Sitter space was given already in the original Gibbons-Hawking 
paper \cite{GibbonsHawking77} and re-proposed in 
\cite{Mottola86}.  Davies \cite{Davies88CQG} has 
considered the event horizon of  FLRW  space and, for GR with a perfect fluid as the source, 
has proved the following theorem: if the cosmological fluid satisfies 
$P+\rho \geq 0$ and $a(t) \rightarrow +\infty$ as $t\rightarrow 
+\infty$, then the area of the event horizon is non-decreasing. 
The entropy of the event horizon is taken to be $S_{EH}=\left( 
\frac{K_Bc^3}{\hbar G} \right) \frac{A_{EH}}{4}$, where $A_{EH}$ 
is its area.  The validity of 
the generalized 2nd law  for certain radiation-filled 
universes was established in \cite{DavisDavies02, 
DavisDaviesLineweaver03}. 

Due to the difficulties with the event horizon one is 
led to consider the apparent horizon instead. Then, 
eq.~(\ref{SAHdot})  tells 
us that, in an expanding universe in Einstein's theory with perfect 
fluid, the apparent horizon area increases except for the 
quantum vacuum equation of state $P=-\rho$ (for which $S_{AH}$  stays 
constant) and for phantom fluids with $P<-\rho$, in which case 
$S_{AH}$ {\em decreases}, adding another element of weirdness  
to the behaviour of phantom matter ({\em e.g.}, 
\cite{phantomthermo} and references therein).

The generalized 2nd law states that the total entropy of 
matter and of the horizon $S_{total}=S_{matter}+S_{AH}$ cannot 
decrease in any physical process,
\be
\delta S=\delta S_{matter}+ \delta S_{AH} \geq 0 \,.
\ee
(We refer here to the apparent horizon, but several authors 
refer instead to the event or particle horizons. The apparent 
horizon is more appropriate since it is a 
quasi-locally defined quantity.)

\section{Entropy of the apparent horizon and scalar-tensor 
gravity}

Black hole thermodynamics has been studied in scalar-tensor and 
other theories of gravity (see \cite{myentropy} for a summary and 
a list of references).  Numerical  studies  show that in the 
intermediate stages of 
collapse of  dust to a black hole in Brans-Dicke gravity,  
the horizon area decreases and the apparent horizon is located 
{\em outside} the event horizon   \cite{ScheelShapiroTeukolsky}. 
Horizon entropy in Brans-Dicke gravity was analyzed by Kang   
\cite{Kang96}, who pointed out that black hole entropy in 
this theory is not  simply one quarter of the horizon area, but 
 rather
\be \label{Kangentropy}
S_{BH}=\frac{1}{4} \int_{\Sigma} d^2x \sqrt{g^{(2)} } \, \phi  
=\frac{\phi A}{4}
\,,
\ee
where $\phi$ is the Brans-Dicke scalar field (assumed to be 
constant 
on the horizon) and $g^{(2)}$ is the 
determinant of the restriction $g_{\mu\nu}^{(2)} \equiv 
g_{\mu\nu} \left.\right|_{\Sigma}$ of the metric 
$g_{\mu\nu}$ to the horizon  $\Sigma$. Naively, this expression 
can be understood by replacing the Newton constant 
$G$ with the effective gravitational coupling 
\be
G_{eff}=\phi^{-1}
\ee
of Brans-Dicke theory; then, the quantity $S_{BH}$ 
is  non-decreasing during black hole collapse \cite{Kang96}. 
Eq.~(\ref{Kangentropy}) has now been derived 
using various procedures \cite{JacobsonKangMyers94, IyerWald94, 
Visser93b}. 
 
As done in \cite{Kang96}, consider the 
Einstein   frame representation of  Brans-Dicke theory given by 
the conformal 
rescaling of the metric
\be
g_{\mu\nu}\longrightarrow \tilde{g}_{\mu\nu} \equiv \Omega^2 \,  
g_{\mu\nu} \,, \;\;\;\; \Omega=\sqrt{ G\phi} \,,
\ee
accompanied by the scalar field redefinition $\phi \rightarrow 
\tilde{\phi}$ with $\tilde{\phi}$ given by
\be
d\tilde{\phi}= \sqrt{ \frac{2\omega +3}{16\pi G}} \,  
\frac{d\phi}{\phi} \;.
\ee
The Brans-Dicke action \cite{BransDicke}
\be
I_{BD}=\int d^4x \, \frac{ \sqrt{-g}}{16\pi}  \left[ \phi R 
-\frac{\omega}{2} \,  g^{\mu\nu} \nabla_{\mu}\phi 
\nabla_{\nu}\phi -V(\phi)  +{\cal  L}^{(m)} \right]
\ee
(where ${\cal L}^{(m)}$ is the matter Lagrangian density) is 
mapped to its  Einstein frame form
\begin{eqnarray}
I_{BD} & = & \int d^4x \,  \sqrt{-\tilde{g}}   \left[  
\frac{ \tilde{R} }{ 16\pi G}  
-\frac{1}{2} \,  \tilde{g}^{\mu\nu} \tilde{\nabla}_{\mu}
\tilde{\phi}  \tilde{\nabla}_{\nu} \tilde{\phi} -U( \tilde{\phi}) 
\right. \nonumber\\
&&\nonumber\\
& &+  \left. \frac{ {\cal  L}^{(m)} }{\left( G\phi \right)^2}  
\right] \,,
\end{eqnarray}
where a tilde denotes Einstein frame quantitites and 
\be
U \left( \tilde{\phi} \right)=\frac{ V(\phi( \tilde{\phi}) )}{ 
\left( G\phi ( \tilde{\phi}) \right)^2} 
\ee
where $\phi=\phi( \tilde{\phi})$.
In the Einstein frame the gravitational coupling is 
constant but matter couples explicitly to the scalar field and  
massive test particles  following geodesics 
of  the Jordan frame $g_{\mu\nu}$  do no longer 
follow  geodesics of $\tilde{g}_{\mu\nu}$ in the Einstein frame. 
Null  geodesics are not changed by the conformal transformation. 
A cosmological or black hole event horizon, being a null surface, 
is also unchanged. The area of an event horizon is not, and the 
change in the entropy formula  $ S_{BH}=\frac{A}{4G} \rightarrow  
\frac{A}{4G_{eff}}=\frac{\phi 
A}{4}  $ is merely the change of the horizon area due to the 
conformal rescaling of $g_{\mu\nu}$. In fact, $ 
\tilde{g}_{\mu\nu}^{(2)} = \Omega^2 \, 
g_{\mu\nu}^{(2)} $ and, since the event horizon is 
not changed, the Einstein frame area is
\be
\tilde{A}=\int_{\Sigma}d^2x \, \sqrt{ \tilde{g}^{(2)}}=
\int_{\Sigma} d^2x \,\Omega^2 \,  \sqrt{ g^{(2)}}=
G \phi \,A 
\ee
assuming that the scalar field is constant on the horizon (if 
this is not true the zeroth law of black  hole thermodynamics 
will not be satisfied). Therefore, the entropy-area relation  
for the event horizon $ \tilde{S}_{BH}=\tilde{A}/4G $  still 
holds in the Einstein frame. 
This is expected on dimensional grounds since, {\em in vacuo}, 
the 
theory reduces to GR  with varying units of 
length $ \tilde{l}_u 
\sim \Omega \, l_u$, time $\tilde{t}_u \sim \Omega \, t_u $, and 
mass $\tilde{m}_u=\Omega^{-1} \, m_u $  (where $t_u, l_u$, and 
$m_u$ are the constant units of time, length, and mass in the 
Jordan frame, respectively). Derived 
units vary accordingly \cite{Dicke}. An area must scale as $A 
\sim \Omega^2 = 
G\phi $ and, in units in which $c=\hbar= 1$ the entropy is 
dimensionless and, therefore,  is not rescaled. As a result, the 
Jordan frame and  Einstein frame entropies coincide 
\cite{Kang96}. 

The equality between black hole entropies in the Jordan and 
Einstein frames is indeed extended   to all theories with action 
$\int d^4x \sqrt{-g} \,  f\left(  g_{\mu\nu}, R_{\mu\nu}, \phi, 
\nabla_{\alpha}\phi  \right) $ which admit an Einstein frame 
representation  \cite{KogaMaeda98}.

As a byproduct of this observation, the Jordan and the Einstein 
frames are  physically equivalent with respect to the entropy of 
the event horizon. A debate on whether these two frames 
are physically equivalent seems to flare up now and again; it 
is pretty well  established that, 
at the classical level, the two frames are simply different 
representations of the same physics \cite{Dicke, Flanagan, 
FaraoniNadeauconfo}. Potential problems 
arise from the representation-dependence of fundamental 
properties of theories of gravity (including the  
Equivalence Principle), but this is not an argumente against 
the equivalence of the two conformal frames: rather, it means 
that such fundamental properties   should, ideally,  be 
reformulated in a  representation-independent way  
(\cite{ThomasStefanoValerio} and references therein).  We will 
not address this problem here.

The classical  equivalence is expected to break down at the 
quantum level; in fact, already in the absence of gravity,  
the quantization of canonically related Hamiltonians  produces 
inequivalent energy spectra  and eigenfunctions 
\cite{inequivalentHamiltonians}. However, it is not clear that 
this happens  at the  semiclassical level for 
conformally related frames \cite{Flanagan}. Black 
hole thermodynamics  is not purely classical (the Planck 
constant $\hbar$ appears in the expressions of the entropy and 
temperature of  black hole and cosmological horizons).  It is 
not insignificant that  the physical equivalence between 
conformal frames holds for the (semiclassical)  entropy of event 
horizons.

Contrary to event horizons, the location of {\em apparent} 
horizons 
(which, in general,  are not null surfaces), is  changed by conformal 
transformations. The problem of relating Einstein frame apparent 
horizons to their Jordan frame counterparts, raised in 
\cite{Alexprevious}, has been solved  in 
\cite{AlexValerio}.

At this point, one may object that there was a logical gap in our 
previous discussion: following common practice, we took the 
formula  $S=A/4G$ for  the (black hole or cosmological) {\em 
event}  horizon  and we used it for  the {\em apparent} horizon. Let us 
consider stationary black  hole or cosmological event horizons 
first. The area  
formula can be derived (in GR or in other theories of gravity) by 
using Wald's Noether charge method \cite{Wald93, IyerWald94, 
IyerWald95, Visser93b, JacobsonKangMyers94} or other methods 
\cite{JacobsonKangMyers94, GibbonsHawking77}.  
As a result, the  usual 
entropy-area relation  remains valid provided  that the 
gravitational coupling $G$ is replaced by the corresponding  
effective  gravitational coupling $G_{eff}$ of the theory, the 
identification of which follows from the  inspection of the 
action or of the  field equations rewritten in the form of 
effective Einstein equations. More rigorously, in 
\cite{Brusteinetal09}  
$G_{eff}$ is identified  by  using the 
matrix  of  coefficients of the kinetic terms 
for metric perturbations \cite{Brusteinetal09}. The metric 
perturbations contributing to the Noether charge in Wald's 
formula are identified with specific 
metric perturbation polarizations associated with fluctuations of 
the  area density on the bifurcation surface $\Sigma$ of the 
horizon (in  $D$ spacetime dimensions, this is the 
$(D-2)$-dimensional spacelike cross-section 
of a Killing horizon on which the Killing field vanishes, and 
coincides with the intersection of the two null hypersurfaces 
comprising this horizon). The horizon entropy is 
\be 
S_{BH}=\frac{A}{4G_{eff}} 
\ee
for a theory described by the action
\be \label{Brusteinaction}
I =\int d^4x \, \sqrt{-g} \, {\cal L} \left( g_{\mu\nu}, 
R_{\alpha\beta\rho\lambda}, 
\nabla_{\sigma} R_{\alpha\beta\rho\lambda}, \phi, 
\nabla_{\alpha}\phi, \, ...\right) \,,
\ee
where $\phi$ is  a gravitational scalar field.  The Noether 
charge is
\be
S=-2\pi \int_{\Sigma} d^2x\, \sqrt{g^{(2)}} \left( \frac{ \delta 
{\cal 
L}}{\delta 
R_{\mu\nu ab}} \right)_{(0)} { \hat{ {\mathbf \epsilon}}} 
_{\mu\nu}  
{\hat{ {\mathbf \epsilon} }}_{\mu\nu}   \,,
\ee
where ${\hat{ {\mathbf \epsilon}}}_{\rho\sigma}   $ is the 
(antisymmetric) binormal vector to the bifurcation 
surface $\Sigma$ (which  satisfies 
$\nabla_{\mu}  \chi_{\nu}={ \hat{  {\mathbf \epsilon} 
}}_{\mu\nu}$ on the bifurcation 
surface $\Sigma$, where $\chi^{\mu}$ is the 
Killing  field vanishing on the horizon) and is  normalized to $
{\hat{ {\mathbf  \epsilon}}}^{ab}   {\hat{ {\mathbf 
\epsilon}}}_{ab}  =-2$. The subscript $(0)$ 
denotes the fact that the quantity in brackets is evaluated on 
solutions of  the equations of motion. The effective 
gravitational coupling is 
then calculated to be \cite{Brusteinetal09}
\be
G_{eff}^{-1}=-2\pi \left( \frac{ \delta {\cal L}}{\delta 
R_{\mu\nu\rho\sigma}} \right)_{(0)} {\hat{ {\mathbf  
\epsilon}}}_{\mu\nu}  
 {\hat{ {\mathbf \epsilon}} }_{\rho\sigma} \,.
\ee

For dynamical black holes, there is no timelike Killing vector to 
provide  a bifurcate Killing horizon. However, it was shown in 
\cite{HaywardMukohyamaAshworth99PLA} that, for apparent horizons,  
the Kodama vector can replace the Killing vector in 
Wald's entropy formula, and the result is one quarter of the area 
in GR (or the corresponding generalization in theories of the 
form~(\ref{Brusteinaction})).  We do not repeat the calculation 
here, but we simply note that the same calculation applies to 
cosmological apparent horizons as well, a point that seems to not 
have 
been noted in the literature on cosmological horizons.

Finally, let us  see how thermodynamics can  restrict the range 
of physical solutions of a  theory of gravity. In Brans-Dicke 
cosmology, consider  the 
exact solution representing a spatially flat universe with  
parameter $\omega=-4/3$, and no matter 
\cite{OHanlonTupper}
\begin{eqnarray} 
a(t) &=& a_0 \exp\left( H \, t \right)  \,,\\
&&\nonumber\\
\phi (t) &=& \phi_0 \exp\left( -3H \, t \right) \,,
\end{eqnarray}
where $a_0$, $\phi_0$, and $H$  are positive constants.
In GR, de Sitter spaces are obtained with constant scalar 
fields but this is not always the case in scalar-tensor 
gravity. Since the entropy of the apparent/event horizon in this 
case is $ S=\phi A_H/4= \phi_0H^{-2} \, \exp\left( -3Ht 
\right)$, it is always decreasing. 
The scalar field plays the role of an effective fluid with 
density and pressure given by the equations of Brans-Dicke 
cosmology which, in the spatially flat case and for vacuum  
and a free Brans-Dicke scalar, reduce to \cite{BransDicke, 
mybook, SalvatoreValerio}
\begin{eqnarray}
&& H^2= \frac{\omega}{6} \, \left( \frac{\dot{\phi} }{\phi} 
\right)^2 
-\frac{H\dot{\phi}}{\phi} \,,\\
&&\nonumber\\
&& \dot{H}= -\frac{\omega}{2} \left( \frac{\dot{\phi}}{\phi} 
\right)^2 +2H \, \frac{\dot{\phi}}{\phi} \,,\\
&&\nonumber\\
&& \ddot{\phi}+3H\dot{\phi}=0 \,.
\end{eqnarray}
In our case, the energy density and pressure of the effective 
fluid are  $\rho_{(\phi)}= -P_{(\phi)} \simeq H^2$. The entropy 
density  of this effective fluid is  
\be
s_{(\phi)}=\frac{P_{(\phi )}+\rho_{(\phi )}}{T}=0 \,;
\ee
therefore,   the  horizon associated with this solution violates 
the generalized 2nd law and  should be regarded as unphysical. 
More generally, the use of entropic considerations to select 
extended theories of gravity, was suggested in 
\cite{BrisceseElizalde08}.

\section{Conclusions}

The apparent horizon suffers from the dependence on the 
spacetime slicing. In FLRW space, it would be unnatural to 
choose a slicing unrelated to the hypersurfaces of spatial 
homogeneity and isotropy and constant comoving time, because 
the latter identify  physical 
comoving observers who see the cosmic microwave background 
homogeneous and isotropic around them (apart from small 
temperature anisotropies of the order $5\cdot 10^{-5}$). 
The problem of the slicing dependence, therefore, does not seem 
so pressing in FLRW spaces, however it is not completely 
eliminated. Nevertheless, apparent horizons seem better 
candidates for thermodynamical  considerations than event or 
particle horizons.

Similar to dynamical black hole horizons, the thermodynamics 
of FLRW cosmological horizons is not 
completely free of problems: the choice of surface gravity 
determines the horizon temperature and there are many 
inequivalent proposals for surface gravity. A natural choice 
of internal energy contained within the apparent (or even event) 
horizon is given by the Misner-Sharp-Hernandez mass, and then it 
is natural to choose the Kodama time and Kodama-Hayward 
surface gravity because the Misner-Sharp-Hernandez mass is 
intimately associated with the Kodama vector as a Noether 
charge. However, doing so, produces a Kodama-Hayward temperature 
which is negative for GR universes with stiff equations of state 
$P>\rho/3$, and it is not obvious that one should give up these 
universes as unphysical. Different choices of temperature would 
produce different forms of the 1st law, corresponding to 
different coefficients for the work term  $dW$ appearing there. 
What is certain, though, is that the apparent (and also the 
event and particle) horizons are not comoving and one should not 
necessarily expect a simple $PdV$ term to appear. 
Perhaps  other choices of quasi-local energy can produce 
consistent forms of the 1st law and be applicable to a larger 
variety of universes; one could turn the argument around and 
use cosmology to help selecting the ``correct'' surface gravity 
also for dynamical black hole horizons (this possibility will be 
the subject of a separate publication).

We have elucidated the causal character of the apparent horizon 
and given a simple criterion for the apparent FLRW  horizon to be 
trapping. This criterion coincides with the one for the  
Kodama-Hayward temperature to be positive-definite, and the 
threshold between trapping and untrapping horizons is a FLRW 
universe filled with conformal matter which, if the weak energy 
condition is assumed, also marks the transition between timelike 
and spacelike nature of the apparent horizon. At the moment, 
we are unable to offer a  simple and consistent physical 
interpretation 
of this fact  and we will refrain from doing so. We have also 
considered the extension of the thermodynamics of cosmological  
horizons to alternative theories of gravity and, within 
Brans-Dicke theory, we have  seen how 
thermodynamical considerations can help judging how physical a 
certain solution can be.

Overall, it appears that the thermodynamics of cosmological 
apparent horizons exhibits features which are not yet fully 
understood. Due to the  extremely simplified nature of FLRW 
spacetime, understanding  these aspects for cosmological apparent 
horizons should be more  fruitful and rapid  than understanding 
the corresponding aspects of black hole dynamical horizons.

\begin{acknowledgments}
This work is supported by the Natural Sciences and  Engineering 
Research  Council of Canada  (NSERC).
\end{acknowledgments}


\end{document}